\documentclass{article}

\begin{document}

\title{An experiential formula for the energy eigenvalues of a particle in a
one-dimension finite-deep square well potential}
\author{Zhi-Ming Zhang\thanks{%
Electronic mail: zzm@sjtu,edu.cn}, Chun-Hua Yuan\thanks{%
Electronic mail: chunhuay@sjtu.edu.cn} \\
Department of Physics, Shanghai Jiao Tong University, \\
Shanghai 200240, China}
\maketitle

\begin{abstract}
We propose an experiential formula for the calculation of the energy
eigenvalues of a particle moving in a one-dimension finite-deep square well
potential after some physical considerations. This formula shows a simple
relation between the energy eigenvalues and the potential papameters, and
can be used to estimate the energy eigenvalues in a very simple way.
\end{abstract}

The issue of a particle moving in a one-dimension finite-deep square well
potential (1D-FDSWP) is a basic problem in quantum mechanics [1]. The usual
approach for finding the energy eigenvalues of this problem is to solve a
set of transcendental equations by numerical calculations. Here in this
paper we propose an experiential formula for the calculation of the energy
eigenvalues of this issue.

Taking the ground state energy $E_{1}=\left( \frac{\pi ^{2}\hbar ^{2}}{%
2ma^{2}}\right) $ of a particle (mass $m$ ) moving in a one-dimension (
direction $x$) infinite-deep (or perfectly rigid walls) square well
potential (width $a$) (1D-IDSWP) as the energy unit, then the energy
eigenvalues for a particle moving in a 1D-IDSWP is

\qquad \qquad \qquad
\begin{equation}
E_{n}=n^{2},n=1,2,3,...
\end{equation}%
and the energy eigenvalues for the particle moving in a 1D-FDSWP (width $a$
and depth $V_{0}$) are usually found by numerically solving the following
transcendental equations%
\begin{equation}
\xi \tan \xi =\eta ,
\end{equation}%
\begin{equation}
\xi ^{2}+\eta ^{2}=\left( \frac{\pi }{2}\right) ^{2}V_{0},
\end{equation}%
for the even parity states, and by solving Eq.(3) together with
\begin{equation}
-\xi \cot \xi =\eta ,
\end{equation}%
for the odd parity states. The values of $\xi $ obtained in this way are
discrete and we denote them by $\xi _{n}$. After finding $\xi _{n}$
numerically the energy eigenvalues can be obtained by
\begin{equation}
E_{n}^{^{\prime }}=\left( \frac{2}{\pi }\right) ^{2}\xi _{n}^{2}.
\end{equation}

Now we make some physical considerations. (a) When $V_{0}\rightarrow \infty $%
, the energy eigenvalues in the case of 1D-FDSWP will reduce to that in the
situation of 1D-IDSWP; (b) In the case of 1D-FDSWP the particle will have
some probability to enter into the potential walls, so the uncertainty $%
\Delta x$ of the particle position is larger than that in the case of
1D-IDSWP. According to the uncertainty principle the uncertainty $\Delta
p_{x}$ of the particle momentum will be smaller than that in the case of
1D-IDSWP, and in turn, the energy eigenvalues in the case of 1D-FDSWP will
be smaller than the corresponding energy eigenvalues in the case of
1D-IDSWP, i.e. $E_{n}^{^{\prime }}<E_{n}$. As $V_{0}$ decreases, $\Delta x$
will increase, while $E_{n}^{^{\prime }}$ decreases (comparing with $E_{n}$%
), so $E_{n}-E_{n}^{^{\prime }}$ becomes larger; (c) For a definite $V_{0}$,
when the energy of the particle increases ($n$ increase), the particle will
enter into the potential wall deeper, so $\Delta x$ increases. Analyzing in
the same way as above, we know that $E_{n}-E_{n}^{^{\prime }}$ will become
larger as $n$ increases. Based on above physical considerations we propose
following experiential formula for the calculation of the energy eigenvalues
of a particle moving in a 1D-FDSWP%
\begin{equation}
E_{n}^{\ast }=n^{2}\left( 1+\alpha \left( \frac{1}{V_{0}}\right) ^{\beta
}\right) ^{-1},
\end{equation}%
By numerical fit we find $\beta =1/2$ and $\alpha \approx 1.3624$, and
therefore%
\begin{equation}
E_{n}^{\ast }=n^{2}\left( 1+\frac{1.3624}{\sqrt{V_{0}}}\right)
^{-1},n=1,2,...,[\sqrt{V_{0}}]+1,
\end{equation}%
where $\left[ A\right] $ denotes the integer part of $A$. The differences
between the energy eigenvalues in a 1D-IDSWP ($E_{n}$) and that in a
1D-FDSWP ($E_{n}^{\ast }$ ) is%
\begin{equation}
\Delta E_{n}\equiv E_{n}-E_{n}^{\ast }=n^{2}\left[ 1-\left( 1+\frac{1.3624}{%
\sqrt{V_{0}}}\right) ^{-1}\right] \approx 1.3624\frac{n^{2}}{\sqrt{V_{0}}},
\end{equation}%
where the sign \textquotedblleft $\approx $ \textquotedblright\ is valid for
$\sqrt{V_{0}}\gg 1.3624$, and this condition is satisfied in most cases.
Eq.(8) clearly shows that $E_{n}^{\ast }<E_{n}$, and their difference $%
\Delta E_{n}\propto \frac{n^{2}}{\sqrt{V_{0}}}$.

In order to compare the results of our experiential formula (7) with that of
the exact one (5) we have done some numerical calculations, among which
Tables I-III are examples. We find that our experiential formula (7) is not
too bad.

Recovering the original energy unit, Eq.(7) can be rewritten as
\begin{equation}
E_{n}^{\ast }=n^{2}\left( 1+1.3624\sqrt{\left( \frac{\pi ^{2}\hbar ^{2}}{%
2ma^{2}}\right) \frac{1}{V_{0}}}\right) ^{-1},n=1,2,...,\left[ \sqrt{%
V_{0}\left( \frac{\pi ^{2}\hbar ^{2}}{2ma^{2}}\right) ^{-1}}\right] +1.
\end{equation}

Eq.(9) establishes a simple relation between the energy eigenvalues and the
potential parameters ($a$ and $V_{0}$) for a particle (mass $m$) moving in a
1D-FDSWP. With this formula we can estimate the energy eigenvalues in a very
simple way.

\bigskip \bigskip

ACKNOWLEDGMENT

This work was supported by the National Natural Science Foundation of China
under Grant no. 60178001.

\bigskip \bigskip

REFERENCES

[1] See for example, L.I.Schiff, Quantum Mechanics, 3rd ed. (McGraw-Hill,
New York, 1968), p.37.

\bigskip \bigskip

Table I. Comparison between $E_{n}^{^{\prime }}$ and $E_{n}^{\ast }$ for $%
V_{0}=15$.

\begin{tabular}{lllll}
$n$ & $1$ & $2$ & $3$ & $4$ \\
$E_{n}^{^{\prime }}$ & $0.7359$ & $2.9209$ & $6.4709$ & $11.1451$ \\
$E_{n}^{\ast }$ & $0.7398$ & $2.9591$ & $6.6579$ & $11.8363$%
\end{tabular}

\bigskip

Table II. Comparison $E_{n}^{^{\prime }}$ and $E_{n}^{\ast }$ for $V_{0}=25$.

\begin{tabular}{lllllll}
$n$ & $1$ & $2$ & $3$ & $4$ & $5$ & $6$ \\
$E_{n}^{^{\prime }}$ & $0.7859$ & $3.1318$ & $6.9965$ & $12.2880$ & $18.7723$
& $25.0010$ \\
$E_{n}^{\ast }$ & $0.7859$ & $3.1435$ & $7.0728$ & $12.5739$ & $19.6467$ & $%
28.2912$%
\end{tabular}

\bigskip

Table III. Comparison $E_{n}^{^{\prime }}$ and $E_{n}^{\ast }$ for $%
V_{0}=64. $

\begin{tabular}{llllllllll}
$n$ & $1$ & $2$ & $3$ & $4$ & $5$ & $6$ & $7$ & $8$ & $9$ \\
$E_{n}^{^{\prime }}$ & $0.8578$ & $3.4275$ & $7.6979$ & $13.6488$ & $21.2441$
& $30.4227$ & $41.0693$ & $52.9030$ & $64.0003$ \\
$E_{n}^{\ast }$ & $0.8545$ & $3.4179$ & $7.6903$ & $13.6717$ & $21.3620$ & $%
30.7613$ & $41.8696$ & $54.6868$ & $69.2130$%
\end{tabular}

\end{document}